\documentclass[a4paper]{article}
\pdfoutput=1 
\usepackage{jcappub} 
\usepackage{amsmath,amssymb,bm,multirow}
\usepackage[dvipsnames]{xcolor}

\usepackage{siunitx}
\usepackage{relsize}
\usepackage{graphicx}
\usepackage[utf8]{inputenc}

\usepackage{mathtools}
\usepackage{algorithmic}

\newcommand{\bk}{\boldsymbol{k}}

\DeclareMathOperator*{\Sum}{\mathlarger{\mathlarger{\mathlarger{\sum}}}}

\usepackage{xcolor}
\usepackage[normalem]{ulem}

\usepackage{aas_macros}

\def\bi#1{\hbox{\boldmath{$#1$}}}

\begin{document}
\author[a,b,c]{Adrian E.~Bayer,}
\author[d,e]{Chirag Modi,}
\author[c,a]{and Simone Ferraro}

\affiliation[a]{Berkeley Center for Cosmological Physics, University of California,
Berkeley, CA 94720}
\affiliation[b]{Department of Physics, University of California,
Berkeley, CA 94720}
\affiliation[c]{Lawrence Berkeley National Laboratory,  1 Cyclotron Road, Berkeley, CA 94720}
\affiliation[d]{Center for Computational Astrophysics, Flatiron Institute, 162 Fifth Avenue, New York, NY 10010}
\affiliation[e]{Center for Computational Mathematics, Flatiron Institute, 162 Fifth Avenue, New York, NY 10010}

\emailAdd{abayer@berkeley.edu}
\emailAdd{cmodi@flatironinstitute.org}
\emailAdd{sferraro@lbl.gov}

\title{
Joint velocity and density reconstruction of the Universe with nonlinear differentiable forward modeling
}


\abstract{
Reconstructing the initial conditions of the Universe from late-time observations has the potential to optimally extract cosmological information. Due to the high dimensionality of the parameter space, a differentiable forward model is needed for convergence, and recent advances have made it possible to perform reconstruction with nonlinear models based on galaxy (or halo) positions. In addition to positions, future surveys will provide measurements of galaxies' peculiar velocities through the kinematic Sunyaev-Zel'dovich effect (kSZ), type Ia supernovae, the fundamental plane relation, and the Tully-Fisher relation. Here we develop the formalism for including halo velocities, in addition to halo positions, to enhance the reconstruction of the initial conditions. 
We show that using velocity information can significantly improve the reconstruction accuracy compared to using only the halo density field. We study this improvement as a function of shot noise, velocity measurement noise, and angle to the line of sight. 
We also show how halo velocity data can be used to improve the reconstruction of the final nonlinear matter overdensity and velocity fields.
We have built our pipeline into the differentiable Particle-Mesh \texttt{FlowPM} package, paving the way to perform field-level cosmological inference with joint velocity and density reconstruction. This is especially useful given the increased ability to measure peculiar velocities in the near future.
}

\maketitle


\section{Introduction}

Reconstructing the initial conditions of the Universe from cosmological data is a pressing task dating back many decades \cite{Hamilton_1997, Tegmark_1997, Bond_1998, Seljak_1998}. Recently, there has been much work to achieve this in the context of galaxy surveys, weak lensing, and the Lyman alpha forest \cite{Seljak_2017, Modi_2018, Modi_2019, Horowitz_2019, Horowitz_2019b, Bohm_2021,Jasche_2013, Jasche_2019, Porqueres_2019}. 
Furthermore, this procedure can be extended to infer cosmological parameters using a field-level approach, which improves constraints compared to a traditional 2-point analysis \cite{Seljak_2017}. This is of particular interest as surveys probe progressively smaller scales in which the effects of nonlinear gravitational evolution moves cosmological information beyond the 2-point statistics. Many summary statistics beyond the power spectrum have been proposed to extract some of this information (see e.g.~\cite{Takada_2004, Sefusatti_2006, Berge_2010, Kayo_2013, Schaan_2014, Liu2015,Liux2015,Kacprzak2016,Shan2018,Martinet2018,liu&madhavacheril2019, Li2019, Kreisch2019, Coulton2019, Sahl_n_2019, Marques2019,ajani2020, Hahn_2020, hahn2020constraining, Dai_2020, Uhlemann_2020, Allys_2020, Gualdi_2020,Harnois-Deraps2020, Arka_2020, Massara_2020, Cheng_2020, Cheng_2021, bayer2021detecting, Kreisch_2021,Bayer_2022_fake, Valogiannis_2021, Valogiannis_2022, Eickenberg_2022}), however, a field-level approach could enable optimal extraction of this information \cite{Seljak_2017}.

While much of the reconstruction literature focuses on using only the galaxy overdensity field as the data, one could consider including additional information in the reconstruction process, such as galaxy peculiar velocities \cite{Prideaux_2022}. This is of particular interest as modern surveys begin to provide accurate measurements of peculiar velocities. Current galaxy peculiar velocity catalogs include the 6dF galaxy survey (using fundamental plane) \cite{Campbell_2014} and Cosmicflows-4 (using the Tully-Fisher relation) \cite{Kourkchi_2020}. 
In addition, the DESI Bright Galaxy Survey \cite{DESI_2016} 
will provide many more measurements of galaxy peculiar velocities using fundamental plane measurements. 

Peculiar velocity information can also be obtained from type Ia supernovae measurements \cite{Riess_2000, Kim_2019}, for example from the DSS survey \cite{Stahl_2021}. This information can be combined with galaxy surveys \cite{Kim_2020} or gravitational waves \cite{Palmese_2021} to understand the nature gravity. Furthermore, upcoming CMB experiments, such as Simons Observatory \cite{SO2018} and CMB-S4 \cite{CMB-S42016} will provide accurate measurements of the kinematic Sunyaev-Zel'dovich (kSZ) effect from which peculiar velocities can be obtained \cite{Deutsch:2017ybc, Smith_2018}.

Adding information from peculiar velocities $v_r$ to the reconstruction framework is expected to greatly reduce reconstruction error on large and intermediate scales. This is because (in linear theory) the reconstruction error from peculiar velocity scales as $k^2$ \cite{Smith_2018, Munchmeyer_2019}, while density reconstruction has an approximately $k$-independent noise. Therefore on sufficiently large scales, depending on the shot noise of the galaxy field, the reconstruction from velocity will have lower noise.
This is of great importance for measuring parameters like primordial non-Gaussianity of the local type which are sensitive to the very large scales and can be measured with reduced sample variance \cite{Seljak:2008xr}. This has been studied analytically in linear theory in \cite{Munchmeyer_2019}, and it is also a sensitive probe of more general models of multi-field inflation as discussed in \cite{Ferraro:2014jba, AnilKumar:2022flx}. Reconstruction from velocities may also help improve the reconstruction of the Baryon Acoustic Oscillations (BAO), especially close to the boundary of the survey \cite{Zhu:2019gzu}.


\begin{figure}
  \begin{center}
    \includegraphics[width=0.6\textwidth]{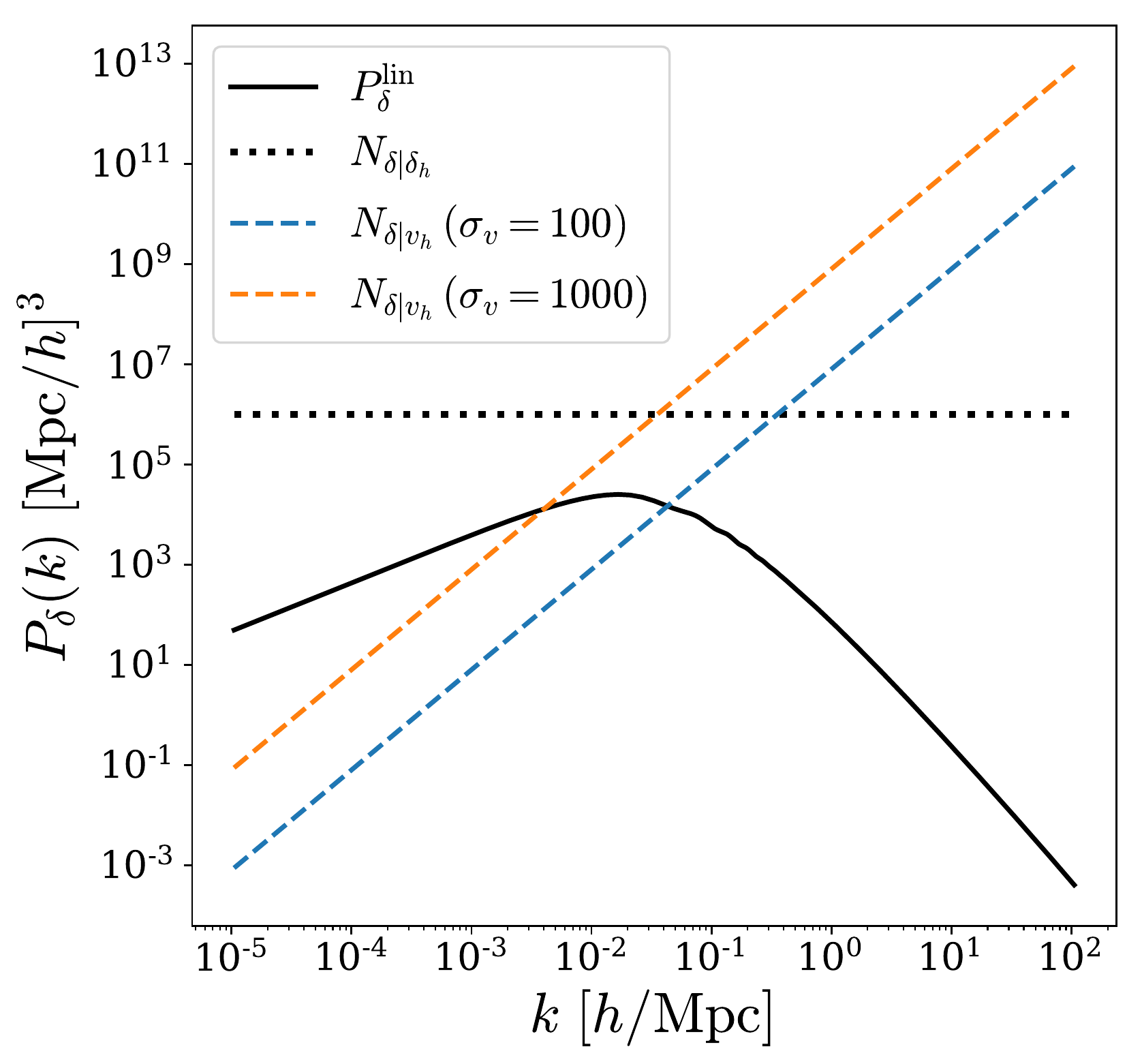}
  \end{center}
  \caption{
  The linear matter power (solid black), shot noise (dotted black), and the error on radial modes of density ($\mu = 1$) given measurements of the line-of-sight velocity with varying measurement error $\sigma_{v_r}$ [in km/s] (dashed colored). The shot noise corresponds to a halo comoving number density of $\bar{n}=10^{-6} (h/$Mpc$)^3$.
  }
  \label{fig:linear}
\end{figure}

The shot noise\footnote{Shot noise arises due to the discrete sampling of objects such as halos and galaxies, 
and is apparent in finite resolution simulations such as our own. 
In principle, if one had perfect knowledge of halo and galaxy formation, and could run a high enough resolution simulation to model it, one could elude shot noise.
While such an endeavour is beyond the scope of this paper, we note that there have been advances in forward modeling progressively smaller scales with progressively higher resolutions to better map the positions of halos from the initial conditions \cite{Sawala_2021}.}, which corresponds to the error in reconstructing the initial field $\delta$ using halo overdensity $\delta_h$ information, is approximately given by
\begin{equation}
    N_{\delta|\delta_h}(k) \approx \frac{1}{\bar{n}},
    \label{eq:shotnoise}
\end{equation}
where $\bar{n}$ is the comoving number density of halos\footnote{We will work with halos, as opposed to galaxies, in the remainder of this paper.}. Note that this is an approximation, and the shot noise can differ from the value above due to the effects of exclusion and non-linear clustering \cite{Baldauf:2013hka}. The effect of this term is to decorrelate the true halo density field from the model halo density field, in turn limiting the accuracy of reconstruction from the density field.

Similarly, it is often the case that peculiar velocities are measured for each halo with independent errors of size $\sigma_{v_r}$ \cite{Smith_2018}. In this case, the noise on the velocity power spectrum is simply given by
\begin{equation}
    N_{v_r|v_{h,r}} \approx \frac{\sigma_{v_r}^2}{\bar{n}}.
    \label{eqn:N_vv}
\end{equation}
We note that $\sigma_{v_r}$ may depend on the redshift of the tracer, with the error growing with distance.

The velocity and density power spectra, in linear theory, are related by the continuity equation:
\begin{equation}
    P_{v_r}(\bk) = \mu^2 \left( \frac{faH}{k} \right)^2 P_\delta(\bk),
    \label{eqn:Pv}
\end{equation}
where $f$ is the growth rate, $a$ is the scale factor, $H$ is Hubble's constant, $\mu$ is the cosine of the angle between $\bk$ and the line of sight (so that $\mu = 1$ corresponds to radial modes), and $P_\delta$ is the linear density power spectrum.
Thus, the error in the density power spectrum given measurements of the velocity is given by
\begin{equation}
    N_{\delta|v_{h,r}} (\bk) = \frac{1}{\mu^2}\left( \frac{k}{faH} \right)^2 N_{v_r|v_{h,r}} \approx \frac{1}{\mu^2} \left( \frac{k}{faH} \right)^2 \frac{\sigma_{v_r}^2}{\bar{n}}.
    \label{eq:rec_noise}
\end{equation}
From the previous equations, we can see that while $N_{\delta|\delta_{h,r}}$ is independent of $k$ (Eqn.~\ref{eq:shotnoise}), $N_{\delta|v_{h,r}}$ scales as $k^2$ (Eqn.~\ref{eq:rec_noise}) and thus we expect the latter to be smaller on sufficiently large scales. 

We illustrate this in Fig.~\ref{fig:linear}, where we show the linear matter power spectrum, together with the error expected from measurement of the halo density field (i.e.~shot noise), as well as the error on density from measurements of the halo velocity for two different velocity measurement errors. It is apparent that for the configuration considered here, we expect reconstruction from velocity to dominate on large scales.

Note that here we focus on reconstruction based on measurements of the \textit{radial} velocities $v_r$ given the more immediate observational prospects. Velocities can also be measured via direct astrometry, e.g.~with GAIA \cite{Nusser_2012}. Moreover, future experiments might also provide measurements of the \textit{transverse} velocity through the ``moving lens'' effect \cite{1983Natur.302..315B, Hotinli:2018yyc}. The same formalism described in this paper applies to reconstruction using transverse velocities or a combination of radial and transverse, potentially with different noise properties.


In recent applications of field-level inference, the 3-dimensional initial density field is typically comprised of many millions of modes (or voxels); this is determined by computational memory and speed, but in principle an arbitrarily fine resolution could be used. Furthermore, the dynamics of the forward model linking the initial field to the final halo field are highly nonlinear.
Reconstruction thus involves performing inference over a highly nonlinear parameter space with millions of dimensions. Knowledge of the derivative of the forward model with respect the initial modes is therefore crucial to efficiently perform the inference.
We use \texttt{FlowPM}\footnote{\url{https://github.com/DifferentiableUniverseInitiative/flowpm}} \cite{Modi_2021}, a differentiable Particle Mesh (PM) simulation code written in TensorFlow. This is entirely differentiable end-to-end, providing derivatives of the forward model, and in turn the data likelihood, with respect to the initial modes and the cosmological parameters. It is also GPU-accelerated which aids the computational efficiency on modern computing nodes\footnote{ For example, on Cori KNL at NERSC, which has 64 CPUs and 4 GPUs per node, one N-body step using GPU-accelerated code is substantially faster. We invite the reader to consult \cite{Modi_2021} for a thorough analysis of the GPU performance of \texttt{FlowPM}, and \cite{Feng2016} for the performance of \texttt{FastPM} (which performs the same computation but is written in C and intended to be run on CPU).}. Furthermore, this technique can readily be combined with machine learning techniques to go beyond the traditional N-body approach, and for example connect galaxies and halos \cite{Modi_2018}.

In this work we apply a fully nonlinear forward model using \texttt{FlowPM} together with a bias model to map matter to halos. 
We assess how reconstruction of the initial field from velocity data compares to reconstruction from density data. We also investigate joint reconstruction of the initial field from a combination of both datasets. We consider the effects of the amount of shot noise, velocity measurement noise, and angle to the line of sight.

A further application one might be interested in is predicting another field given the reconstructed initial field. 
This can simply be achieved by running high fidelity forward models using the reconstructed initial density field. For example, to reconstruct the final matter velocity field, one can run a forward model for the final matter velocity field on the reconstructed initial field.
In the case of reconstructing the final matter velocity field, attempts have been made using various approximations, such as assuming the velocity follows a scalar potential and linear approximations \cite{Bertschinger_1989, Dekel_1999, Willick_1997, Willick_1998, Hoffman_1991, Sorce_2016, Zaroubi_1999, Freudling_1999, Courtois_2012, Tully_2008, Tully_2014, Hoffman_2015, Leclercq_2015, Lavaux_2016, Leclercq_2017}.
Applications to kSZ measurements are also considered in \cite{Nguyen:2020yuc}.
More recently \cite{Prideaux_2022} used the BORG algorithm \cite{Jasche_2013, Jasche_2019} 
to perform reconstruction with peculiar velocity data using a Lagrangian perturbation theory (LPT) forward model on dark matter, finding it to outperform linear reconstruction. To improve upon this, in this work we extend to using a fully nonlinear forward model and consider performing reconstruction with joint velocity and density data.

The structure of the paper is as follows. In Section \ref{sec:method} we review the Bayesian procedure of initial mode reconstruction, and discuss how it can be used to perform field-level inference. We also discuss the datasets and forward model employed in this paper. In Section \ref{sec:results} we show the quality of reconstruction for both, the initial (linear) and final (nonlinear) matter density fields, as well as the matter velocity field. We conclude in Section \ref{sec:conclusions}.



\section{Method}
\label{sec:method}

In this section we review the Bayesian methodology of field-level inference. We start in the context of reconstruction from density data, and then describe how velocity can be included. We then discuss the data and forward model employed in this work. For the sake of clarity we will present the theory both for reconstruction and cosmological parameter inference, even though we will not perform cosmological parameter inference in this particular work. 

\subsection{Field-level inference}
\label{sec:fieldlevel}

Bayes theorem states that posterior distribution of parameters $\theta$ given data $d$ is given by
\begin{equation}
    p(\theta|d)=\frac{p(d|\theta)p(\theta)}{p(d)},
    \label{eqn:bayes}
\end{equation}
where $p(d|\theta)$ is the likelihood, $p(\theta)$ is the prior, and $p(d)$ is the evidence.
In the cosmological context $\theta$ might refer to cosmological parameters which we wish to estimate, and $d$ can refer to the galaxy overdensity field or peculiar velocity data.
In principle one would like to estimate the full posterior distribution of $\theta$, but a more tractable, approximate, alternative is to estimate the best-fit value, i.e.~the maximum a posteriori (MAP) value, of $\theta$, denoted $\hat{\theta}$, and the width of the posterior around the maximum to quantify the uncertainty of the estimate.

In addition to the parameters we wish to infer, $\theta$, there will typically also be nuisance parameters which we wish to marginalize out of the problem, denoted $z$. In cosmology, this corresponds primarily to the initial conditions i.e.~the initial density distribution of the Universe. Then, the marginalized likelihood required to evaluate Eqn.~\ref{eqn:bayes} is computed by integrating out the nuisance parameters from the joint likelihood $p(d|\theta,z)$  as follows,
\begin{equation}
    p(d|\theta) 
    = \int dz~ p(z,d|\theta)
    = \int dz~ p(d|\theta,z) p(z|\theta),
    \label{eqn:marg}
\end{equation}
where $p(z|\theta)$ is the prior of $z$ conditioned on $\theta$.

Thus, to compute the posterior, $p(\theta|d)$, Eqns~\ref{eqn:bayes} and \ref{eqn:marg} show that there are 3 ingredients required: the joint likelihood, $p(d|\theta,z)$, the prior of $z$ conditioned on $\theta$, $p(z|\theta)$, and the prior on $\theta$, $p(\theta)$.\footnote{Note that the evidence, $p(d)$, is a constant and can be dropped when one is only interested in finding the maximum or width of the posterior.} We will now discuss each ingredient separately.

\subsubsection{Likelihood}

In this work we consider data corresponding to the halo overdensity field $\delta$ and peculiar velocity along the line of sight $v$, thus $d=\{\delta, v\}$\footnote{From now on, we use the shorthand $v$ to denote the peculiar velocity. While we only consider peculiar velocities in this work, we note that our formalism is general and can be applied to any component of the velocity field.}. The halo overdensity field data corresponds to a 3D mesh containing the value of the overdensity field in each voxel. For clarity of notation, the overdensity field $\delta$ can be thought of as a vector consisting of each pixel in the map.
On the other hand the peculiar velocity data corresponds to the peculiar velocity of each halo. We thus note that the overdensity is considered at the field level, while the velocity data is considered at the object level.

To reconstruct any latent field, one must make use of a forward model of the data $f(\theta,z,...)$ which in general depends on the parameters $\theta$ and $z$. This typically corresponds to a perturbative model or N-body simulation. 
There will inevitably be some error in the forward model, as well as some noise in the data. In this analysis we assume these to be Gaussian and uncorrelated, with variance
\begin{equation}
    \sigma^2=\sigma^2_{\rm model} + \sigma^2_{\rm data}.
    \label{eqn:sigma}
\end{equation}
Under this assumption, the negative log likelihood is given by the chi-squared difference between the data and a forward model.
For halo overdensity data alone the likelihood is thus given by
\begin{equation}
   -2 \log p(\delta|\theta,z) = \sum_{\bi{k}} \frac{|\tilde{\delta}(\bi{k}) - f_{\tilde{\delta}}(\bi{k};\theta,z)|^2}{\sigma_{\tilde{\delta}}^2(\bi{k})},
   \label{eqn:L_delta}
\end{equation}
where the sum is performed over all modes \bi{k}, and
the $\delta$ subscript denotes this is the forward model and error for the overdensity data.
Note that the density field is evaluated in Fourier space (denoted by the $\sim$), as the model error is typically $k$ dependent. Eqn.~\ref{eqn:L_delta} corresponds to approximating the likelihood to be Gaussian; this assumption works well on large and intermediate scales, but a more sophisticated likelihood would be required to accurately describe small scales, and also to account for observational systematics such as masking, light cones, luminosity dependence, and depth modulation.

Analogously to Eqn.~\ref{eqn:L_delta}, for velocity data alone we have
\begin{equation}
   -2 \log p(v|\theta,z) = \sum_i \frac{[v_i - f_v(\bi{x}_i;\theta,z)]^2}{\sigma_v^2},
   \label{eqn:L_vel}
\end{equation}
where the sum is over all velocity tracers, labeled by $i$. The forward model for the velocity of the $i^{\rm th}$ halo, $f_v(\bi{x}_i;\theta,z)$, depends on the position of the halo $\bi{x}_i$, as will be described in Sec.~\ref{sec:vel_fwd}. We note that a more sophisticated likelihood may be required when analyzing data, for example depending on the technique used to extract the signal (e.g.~matched filter), and including a tracer-dependent error (here $\sigma_v$ is the same for all tracers).

While we will consider both density and velocity data individually in this paper, we will also consider the effects of combining density and velocity data. In such a case, and under the assumption of independence, the log likelihood is simply given by the sum of Eqns.~\ref{eqn:L_delta} and \ref{eqn:L_vel},
\begin{align}
    -2 \log p(\delta,v|\theta,z) 
    &=-2 \log p(\delta|\theta,z)-2 \log p(v|\theta,z) \nonumber\\
    &= \sum_{\bi{k}} \frac{|\tilde{\delta}(\bi{k}) - f_{\tilde{\delta}}(\bi{k};\theta,z)|^2}{\sigma_{\tilde{\delta}}^2(\bi{k})} + \sum_i \frac{[v_i - f_v(\bi{x}_i;\theta,z)]^2}{\sigma_v^2}.
    \label{eqn:L_joint}
\end{align}

\subsubsection{Priors}

There are two priors to consider, the prior of $z$ conditioned on $\theta$, $p(z|\theta)$, and the prior on $\theta$, $p(\theta)$.
In the context of density reconstruction, the nuisance parameters $z$ refer to the initial overdensity field modes, and $\theta$ corresponds to the cosmological parameters.
The prior on $\theta$ will typically be motivated by previous measurements of the cosmological parameters from experiments such as Planck \cite{planck2018}. 
In our analysis we will not do inference on $\theta$ and fix $\theta$ at it's true value, thus this prior term is not relevant.
Based on inflationary theory, which has been verified by Planck, the prior on the initial modes $z$ is taken to be Gaussian with mean 0 and variance proportional to the power spectrum. Hence the prior of $z$ conditioned on $\theta$ is given by
\begin{equation}
    -2 \log p(z|\theta) = \sum_{\bi{k}} \frac{|\tilde{z}(\bi{k})|^2}{P(k;\theta)},
    \label{eqn:prior}
\end{equation}
where $P(k;\theta)$ is the power spectrum of the initial modes,
and depends on the cosmological parameters $\theta$. Note that the initial modes are written in Fourier space, and can thus be complex.

\subsubsection{Posterior}

Adding Eqns \ref{eqn:L_joint} and \ref{eqn:prior} gives the posterior of $z$ given $\theta$ for joint density and velocity inference,
\begin{equation}
    -2 \log p(z,d=\{\delta,v\}|\theta)
    = \sum_{\bi{k}} \left\{ \frac{|\tilde{\delta}(\bi{k}) - f_{\tilde{\delta}}(\bi{k};\theta,z)|^2}{\sigma_{\tilde{\delta}}^2(\bi{k})}
    + \frac{|\tilde{z}(\bi{k})|^2}{P(k;\theta)} \right\} + \sum_i \frac{[v_i - f_v(\bi{x}_i;\theta,z)]^2}{\sigma_v^2}.
    \label{eqn:posterior}
\end{equation}
Note that it is $z$, not $\tilde{z}$, that appears on the left hand side as we perform the inference on the initial field in configuration space, enforcing the physical constraint that $z$ is a real field. 
To compute the posterior on $\theta$ from Eqn.~\ref{eqn:bayes}, one must compute the integral over initial modes from Eqn.~\ref{eqn:marg}. This can be done using traditional sampling methods, such as HMC as in the BORG method \cite{Jasche_2013}, or via optimization with the Laplace approximation or other approximation schemes \cite{Seljak_2017,Millea_2022}. In this work we are motivated by the optimization approach, whereby the first step to performing the marginalization integral is to find the value of $z$ which maximizes $p(z,d|\theta)$. In this approach one will first iterate to find the MAP $\theta$, and then find the MAP $z$ to perform the inference.
We refer the reader to \cite{Seljak_2017} for further details, but we mention this to note that finding the MAP $z$ is not only interesting from the perspective of reconstructing the initial modes, but also in terms of parameter inference.
In this work we will focus on computing $\hat{z}\equiv\max_z p(z,d|\theta)$, in which the cosmological parameters $\theta$ are fixed at their true values.
We will explore performing inference on $\theta$ in future work.
We will additionally be interested in the quality of the reconstructed final  matter density and velocity fields, which we compute by running the forward model using $\hat{z}$ as the initial conditions. We note that these results could be biased as the forward modeled MAP initial field is not necessarily the MAP final field -- the MAP does not generally commute with the nonlinear forward model. A more thorough, unbiased, approach would be to go beyond the MAP and obtain the full posterior, but that is beyond the scope of this work. We note that one could also use more accurate forward models to obtain the final fields from the inferred initial field to improve the small scale agreement (e.g.~\cite{Ata_2022}).

\subsection{Data}
\label{sec:data}

We consider simulated halo field as data observables. To generate this data, we use \texttt{FastPM}  \cite{Feng2016,Bayer_2021_fastpm} with $N_{\rm cdm} = 1024^3$ CDM particles, and a $N_{\rm grid} = 2048^3$ resolution force grid. 
We use the following cosmological parameters: $\Omega_m = 0.3175$, 
$\Omega_b=0.049$, $h=0.6711$, $n_s=0.9624$, 
$\sigma_8=0.834$, and $M_\nu=0$. We begin the simulation at redshift of $9$ and use 20 steps to evolve to redshift $0$. The halo catalog is computed using the Friends-of Friends (FoF) algorithm with a linking length of 0.2. The halo positions $\bi{x}_h^{\rm data}$ and velocities $v_h^{\rm data}$ are computed at the halo center-of-mass using \texttt{nbodykit} \cite{Hand_2018}. The halo overdensity field is computed using the cloud-in-cell (CIC) method. We do not consider the effect of redshift space distortions in the main text, as they have little effect on the reconstruction accuracy for halo models on the scales considered in this work. We explicitly show the effect of redshift space distortions on reconstruction in Appendix \ref{app:rsd}. 

Our fiducial data considers a box of side length $L=4\,{\rm Gpc}/h$. We select the 67,000 most massive halos, corresponding to a number density of $\bar{n}\simeq10^{-6}\, (h/ {\rm Mpc})^3$, a minimum halo mass of $M_{\rm min}\simeq6.8 \times 10^{14} M_\odot / h$, and a bias of $b_1\simeq4.0$. This data has a high shot noise to illustrate the benefits of velocity reconstruction.

We inject white noise into the velocity data with standard deviation $\sigma_{v,{\rm data}} = 300 \,{\rm km/s}$. We inject no noise into the density field $\sigma_{\delta,{\rm data}}=0$ (although there is still the natural Poisson shot noise due to considering discrete tracers) to understand how helpful velocity data is in this limiting case. Throughout the results we will consider the effect of perturbing individual components of this fiducial setup.

We will additionally consider a lower shot-noise example consisting of a $L=400\,{\rm Mpc}/h$ box with the 10,000 most massive halos, corresponding to a number density of $\bar{n}\simeq1.6\times10^{-4}\, (h/ {\rm Mpc})^3$, a minimum halo mass of $M_{\rm min}\simeq3.0\times 10^{13} M_\odot / h$, and a bias of $b_1\simeq0.67$. This example allows studying effects on smaller, nonlinear, scales.

\subsection{Forward models and errors}

In this subsection we describe our forward models for both the halo overdensity, $f_{\tilde{\delta}}(\bi{k};\theta,z)$, and the peculiar velocity, $f_v(\bi{x}_i;\theta,z)$.
We will also discuss the model errors.

\subsubsection{Halo overdensity field forward model}

\begin{figure}
  \begin{center}
    \includegraphics[width=\textwidth]{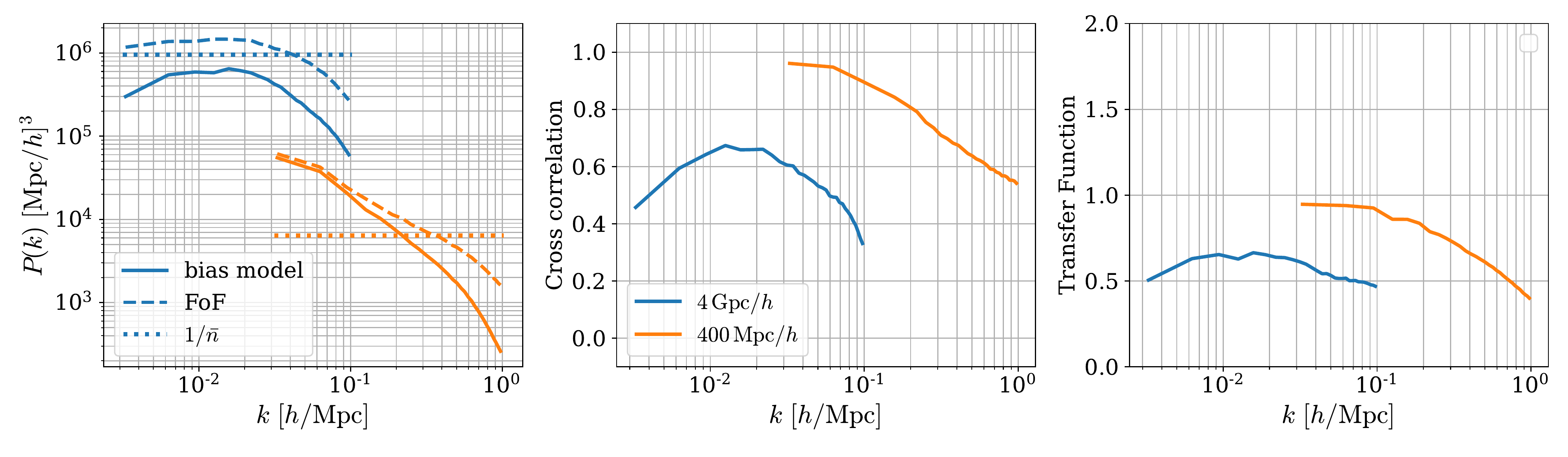}
  \end{center}
  \caption{Comparison of the best fit bias model to the true FoF halo data. We plot for the fiducial $4\,{\rm Gph}/h$ box (blue) when the signal-to-noise is low, and a $400\,{\rm Mph}/h$ box with higher signal-to-noise (orange). Left: bias model (solid), true FoF (dashed) power spectra, and Poisson shot noise (dotted). Middle: cross correlation between model and truth. Right: transfer function between model and truth.}
  \label{fig:fof_bias_fit}
\end{figure}

We first compute the matter overdensity field using \texttt{FlowPM} \cite{Modi_2021} (a TensorFlow version of the \texttt{FastPM} simulation which was used to generate the data). For the matter-halo connection we use a linear LPT effective field theory model, namely
\begin{equation}
    f_{\tilde{\delta}}(\bi{k}) = \tilde{\delta}_h^{\rm model} (\bi{k}) = \int d^3\bi{q} \left[ 1 + b_1 \delta_1(\bi{q}) \right] e^{ - i\bi{k} \cdot \left( \bi{q} + \bi{\psi}(\bi{q}) \right)},
    \label{eqn:bias}
\end{equation}
where $k$ is the wavenumber, $q$ are the grid coordinates, $b_1$ is the linear bias parameter, $\delta_1$ is the linear matter overdensity field (which corresponds to $z$ in the Sec.~\ref{sec:fieldlevel} discussion), and $\psi(q)$ is the Lagrangian displacement field. The cosmological parameter $\theta$ dependence enters via $\delta_1$ and $\psi$, but we drop this from the notation to avoid clutter.

We compute $\psi$ with \texttt{FlowPM} using 5 steps between redshift $9$ and $0$. We use $N_{\rm cdm} = 128^3$ matter particles, and force grid with resolution $N_{\rm grid} = 128^3$. Note the differences in the forward model and the data generation described in Sec.~\ref{sec:data}.

We define the model error as the `stochastic' or `shot noise' term, 
\begin{equation}
    \epsilon (\bi{k}) = \tilde{\delta}_h^{\rm model}(\bi{k}) - \tilde{\delta}_h^{\rm data}(\bi{k}).
    \label{eqn:eps}
\end{equation}
The variance on the overdensity model is thus given by the error power spectrum, defined as 
\begin{equation}
    \sigma^2_{\tilde{\delta}, {\rm model}} (k) = P_{\rm err} (k) = \frac{1}{N_{\rm modes}(k)} \sum_{\bi{k}:|\bi{k}|=k} |\epsilon(\bi{k})|^2,
\end{equation}
where $N_{\rm modes}(k)$ in the number of modes in the $k$ bin.

The integrand in the square brackets of Eqn.~\ref{eqn:bias} is equal to the Lagrangian field at $q$, namely $\delta_h^L(q) = 1+b_1 \delta_1(q)$. One could extend this relation to include higher order bias terms (see e.g.~\cite{Schmittfull_2018, Schmidt_2021a, Schmidt_2021b}), however for the number density of tracers considered here, we did not find improvement from using higher order terms and thus a linear bias model is sufficient for the purpose of this work. We note that the choice of model can bias the reconstruction in various ways (see \cite{Nguyen_2021} for a detailed review).

We fit the bias parameter $b_1$ before performing reconstruction by minimizing $|\epsilon(\bi{k})|^2$, from Eqn.~\ref{eqn:eps}, as in \cite{Schmittfull_2018}. Note we only fit using scales with $k \leq N_{\rm cdm}/L$; this choice is somewhat arbitrary, but is chosen because the bias model breaks down at high $k$ (see e.g.~\cite{Schmittfull_2018,Nguyen_2021}). We then fix the bias parameter to this best fit value throughout the analysis -- in a full analysis one would infer the bias parameters, along with the cosmological parameters, while performing the reconstruction.
We show the best fit model for our fiducial $4 \,{\rm Gpc}/h$ and $400 \,{\rm Mpc}/h$ setups (described in Section \ref{sec:data}) in Fig.~\ref{fig:fof_bias_fit}. It can be seen that the cross correlation and transfer function between the model and truth is considerably less than unity in regions of high shot noise. This occurs on all scales for the bigger box, while only on small scales for the small box. This is expected since we are trying to fit a discrete tracer field with a continuous bias model field, which limits the ability of density-based reconstruction. Note that one could also consider a neural network forward model for the halo overdensity field \cite{Modi_2018} to model individual discrete objects, but we do not pursue that here.
The velocity-based reconstruction however does not suffer from this noise.

\subsubsection{Peculiar velocity forward model}
\label{sec:vel_fwd}

\begin{figure}
  \begin{center}
    \includegraphics[width=.317\textwidth]{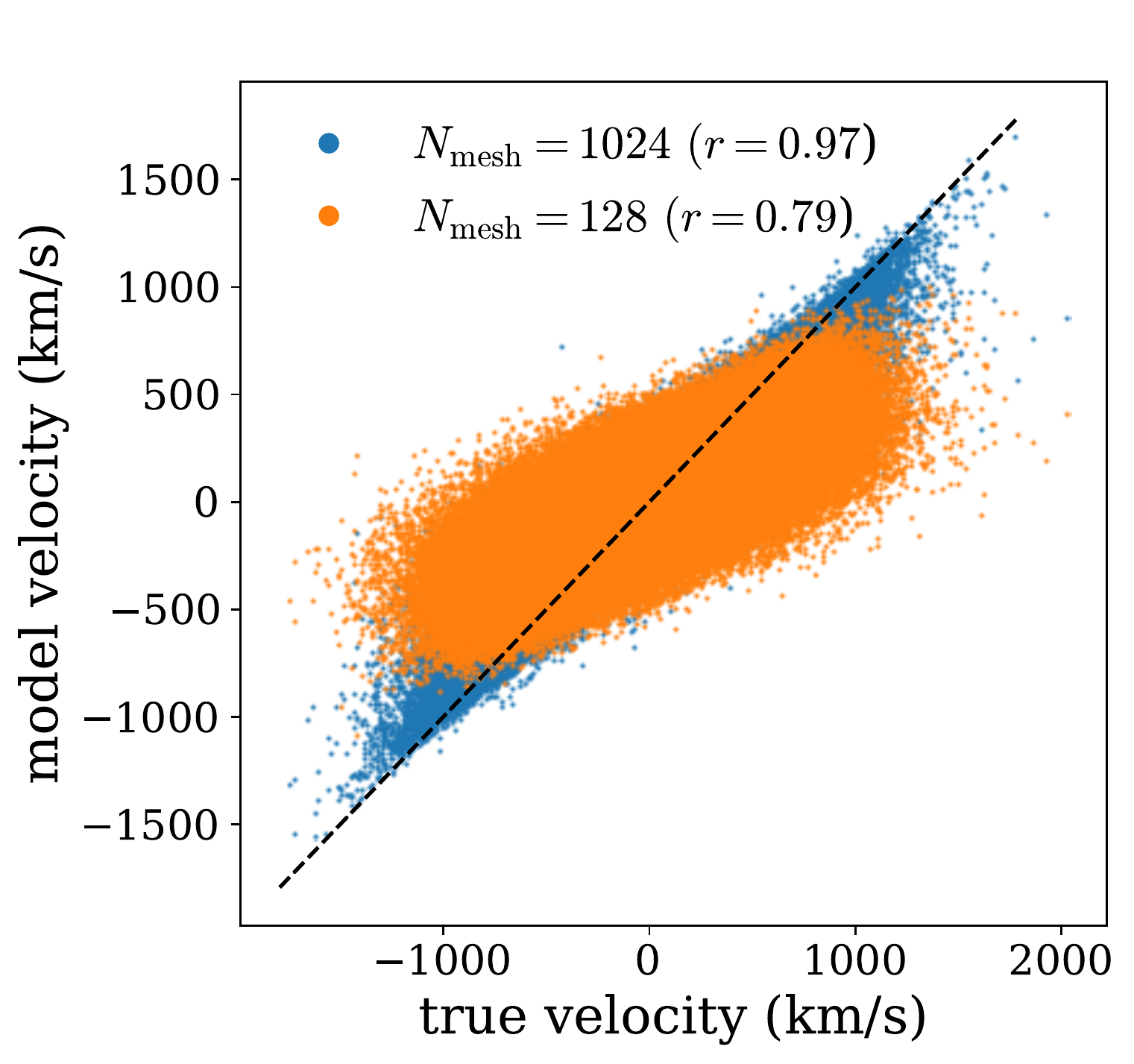}
    \includegraphics[width=.31\textwidth]{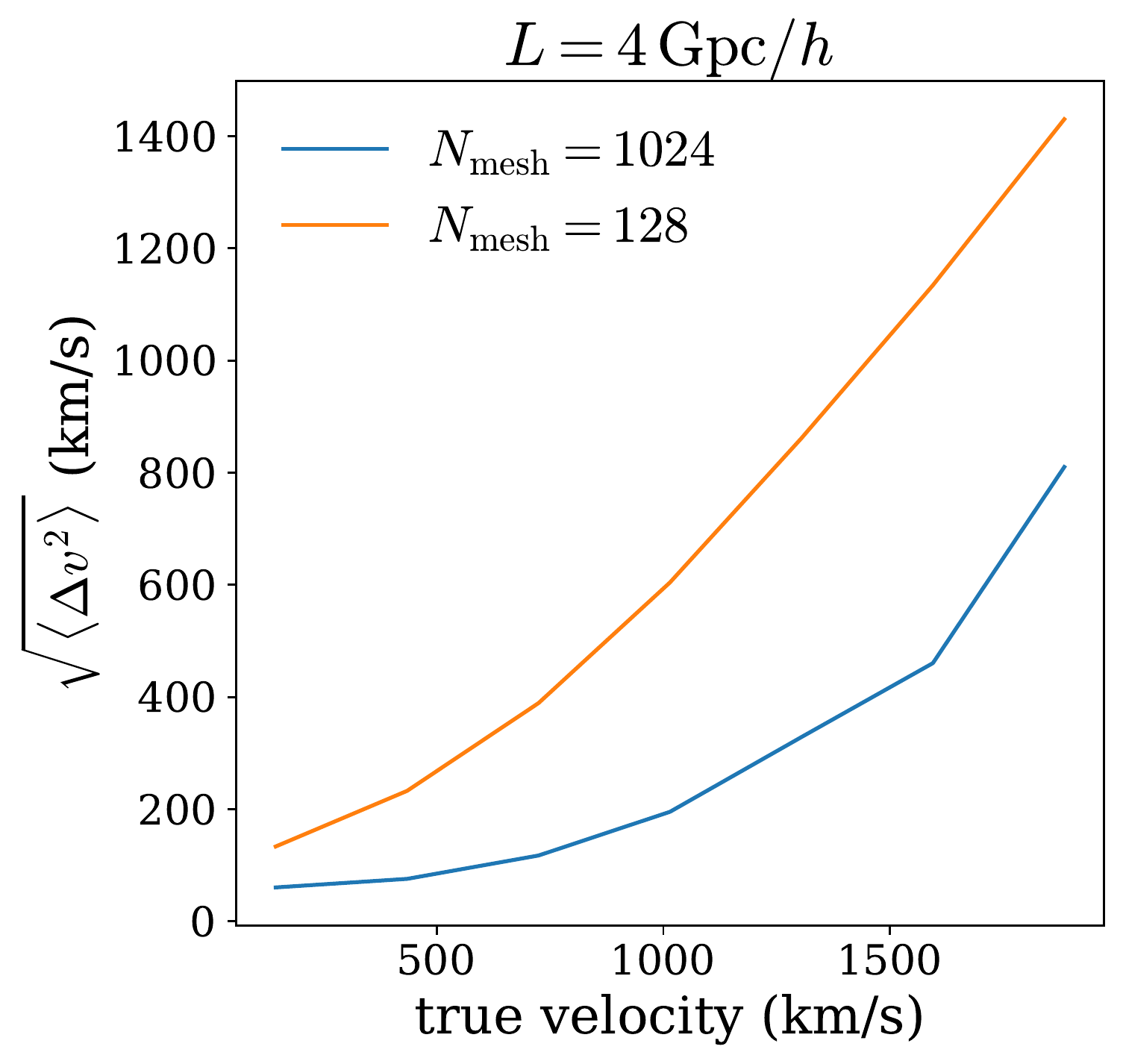}
    \includegraphics[width=.295\textwidth]{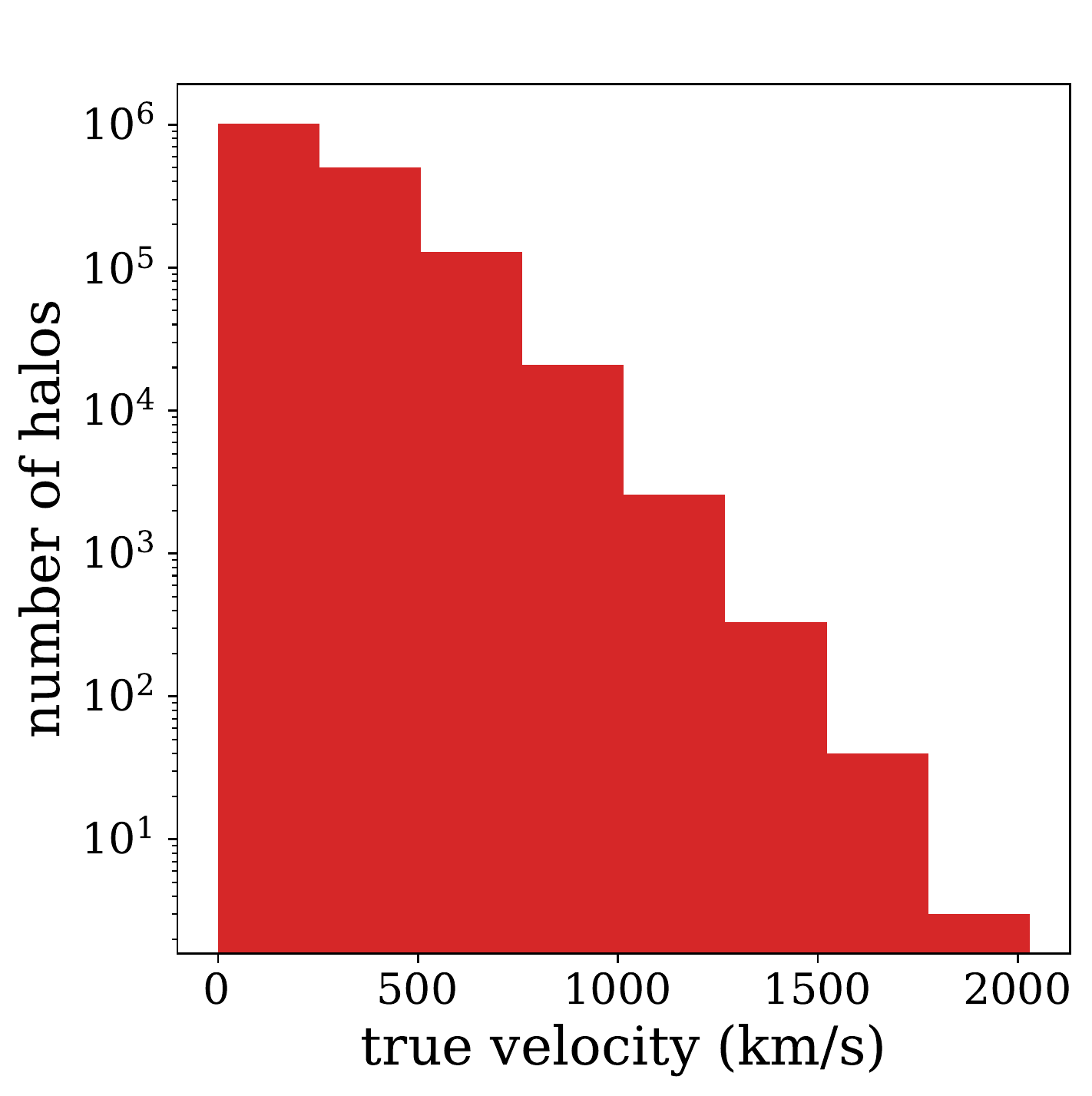}
  \end{center}
  \caption{Velocity information for fiducial ($4\,{\rm Gpc}/h$) setup. Left: scatter plot of the model velocity compared to the true (data) velocity for an $N_{\rm mesh} = 1024$ (blue) and $N_{\rm mesh} = 128$ (orange) forward model. Middle: the corresponding root squared difference between the truth and model velocities. Right: halo abundances as a function of velocity.}
  \label{fig:verror}
\end{figure}

For the velocity forward model we assume an unbiased forward model, i.e. we ignore halo velocity bias terms which should vanish on the large scales considered here due to the equivalence principle \cite{Baldauf:2014fza}. We thus assume that the halo velocity directly traces the underlying matter velocity field. This is not a complete description because, for example, halos may be in multistreaming regions (see e.g.~\cite{Ciecielg_2003, Colombi_2007, Lavaux_2016}), but such effects are beyond the scope of this work. The matter velocity field can be computed using the final matter particle position and velocities from \texttt{FlowPM} as follows. First we define the momentum field at position $\bi{x}$ as,
\begin{equation}
    \mathcal{V}(\bi{x}) = [1+\delta(\bi{x})]v(\bi{x}).
\end{equation}
This is evaluated using CIC interpolation to paint the overdensity field $\delta$ weighted appropriately by the velocities $v$. 
As noted above, in this work we assume that the velocity is only measurable along the line of sight, thus, we simply use $v$ to denote the velocity projected along the line of sight. 
%
Given data corresponding to halo positions $\bi{x}_h^{\rm data}$ and peculiar velocities $v_h^{\rm data}$, the model velocity is computed as 
\begin{equation}
    f_v(\bi{x}_h^{\rm data}) \equiv v_h^{\rm model} = \frac{\mathcal{V}(\bi{x}_h^{\rm data})}{1+\delta(\bi{x}_h^{\rm data})}.
    \label{eqn:vmodel}
\end{equation}

The error in this forward model has a strong dependence on the mesh resolution used in the CIC assignment scheme. Fig.~\ref{fig:verror} shows the scatter between the true halo velocity and model velocity (left), and the corresponding error variance on the model velocity as a function of the true velocity (middle). The model velocity corresponds to Eqn.~\ref{eqn:vmodel} using $N_{\rm mesh}=1024$ (blue) and $N_{\rm mesh}=128$ (orange) for CIC. 
It can be seen that the model error is larger when using a coarser mesh. For $N_{\rm mesh}=128$, the error approximately scales as $0.7v_h^{\rm true}$, i.e. it is an 70\% effect. On the other hand, for $N_{\rm mesh}=1024$ the error is almost negligible for low velocity halos, and only a $\sim 20\%$ effect for the fastest halos. In this work, for tractable reconstruction, we use a $N_{\rm mesh}=128$ for the forward model. Since the majority of halos have low velocities (Fig.~\ref{fig:verror} right panel) this is able to provide a sufficient quality of reconstruction for this work,
however one can expect the results to improve as we use higher resolution forward models. We thus use an interpolated form of the orange line for velocity model error, $\sigma_{v, {\rm model}}$.

\subsection{Optimization}
\label{sec:opt}
Given the data and forward model, we maximize the posterior to get the MAP estimate of the initial field. 
Since the parameter space consists of $128^3 \approx 2$ million dimensions, we need to use optimization algorithms which make use of the gradient information that is readily provided by our differentiable PM code. In this work, we use the LBFGS-B algorithm \cite{Byrd_1995:LBFGS} which uses gradients at each step, and additionally keeps track of them over the trajectory to approximate the Hessian with a low memory cost. 
As there is much noise, and many more modes to be fitted on small scales than large scales, we anneal the posterior as we optimize to iteratively fit the modes up to a give scales $k<k_{\rm iter}$. Mathematically, we multiply the density term in the loss from Eqn.~\ref{eqn:L_joint} with a step function $A(k-k_{\rm iter})$ as follows,
\begin{align}
    -2 \log p_{\rm iter}(z,d=\{\delta,v\}|\theta)
    =& \sum_{\bi{k}} \left\{ A(k-k_{\rm iter}) \frac{|\tilde{\delta}(\bi{k}) - f_{\tilde{\delta}}(\bi{k};\theta,z)|^2}{\sigma_{\tilde{\delta}}^2(\bi{k})}
    + \frac{|\tilde{z}(\bi{k})|^2}{P(k;\theta)} \right\} \nonumber\\
    &+ \sum_i \frac{[v_i - f_v(\bi{x}_i;\theta,z)]^2}{\sigma_v^2},
    \label{eqn:posterior_anneal}
\end{align}
where $A(k) = 1$ if $k\leq0$ and $0$ if $k>0$. We iteratively increase $k_{\rm iter}$ in steps of the fundamental frequency, $k_F = 2 \pi / L$, up to some maximum value $k_{\rm max}$ beyond which convergence breaks down. In this work we use $k_{\rm max} = 16 k_F$. We note that the cutoff for scales smaller than $k_{\rm max}$ could affect the quality of reconstruction on scales larger than the cutoff \cite{Nguyen_2021}, but such effects are sufficiently small to not affect the conclusions of this work. A similar annealing approach has been applied and studied in \cite{Modi_2018}.

To ensure the optimizer has sufficiently converged we average over 5 datasets, each starting at a different initial guess for the initial modes. We refer the reader to \cite{Feng_2018} for a thorough analysis of the posterior surface and how the starting position of the optimizer can affect the reconstruction, but such effects are sufficiently small to not affect the conclusions of this work.

\section{Results}
\label{sec:results}


\begin{figure}
\vspace{-1cm}
  \begin{center}
    \includegraphics[width=\textwidth]{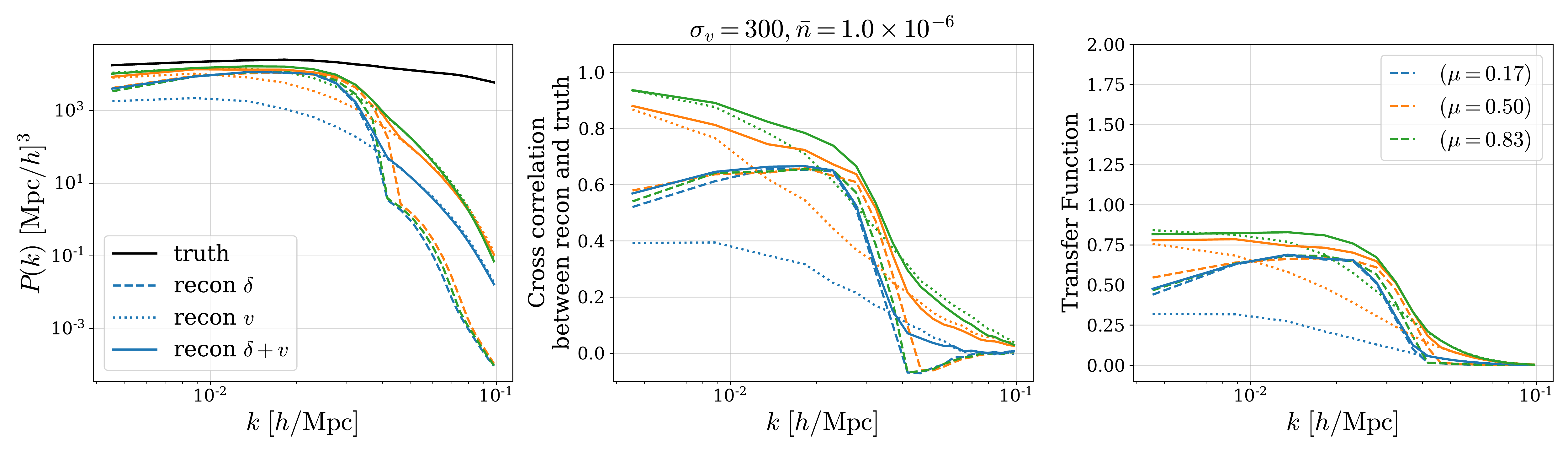}
    \includegraphics[width=\textwidth]{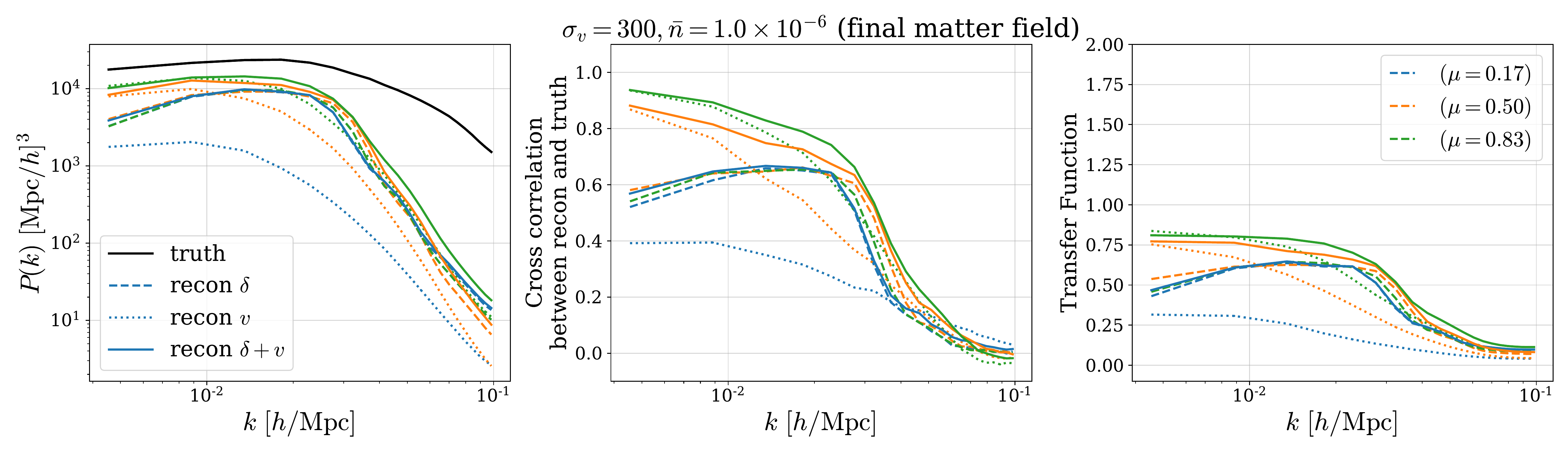}
    \includegraphics[width=\textwidth]{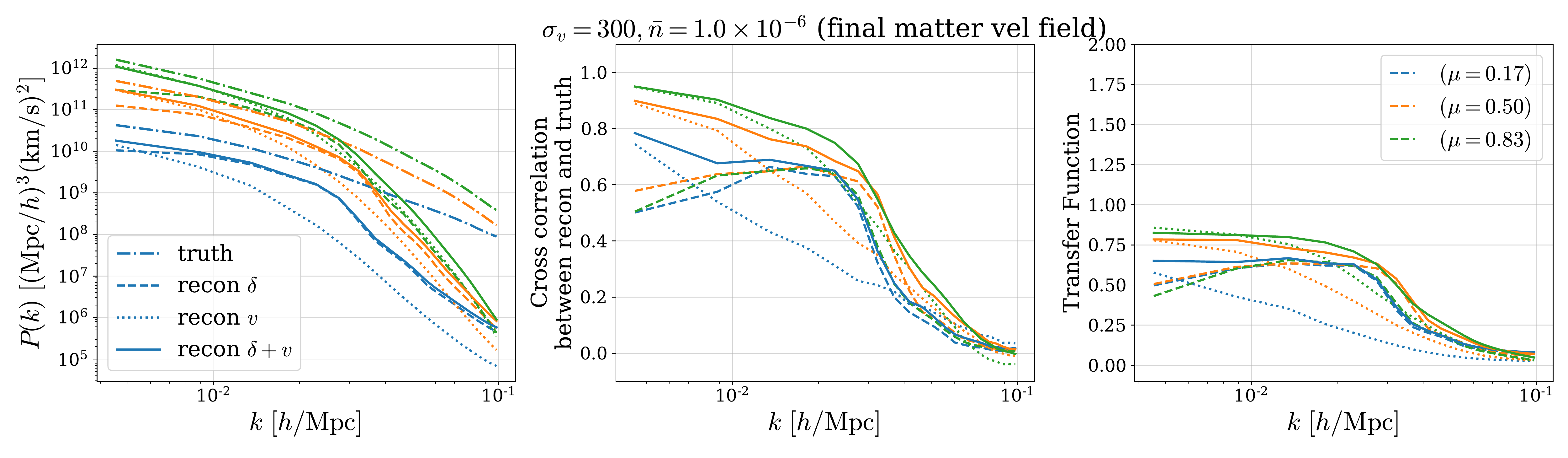}
  \end{center}
  \caption{Reconstruction of the initial linear matter field, final matter field, final matter velocity field (top to bottom) for fiducial setup:  $4\, {\rm Gpc}/h$, $\bar{n}=10^{-6}\, (h/ {\rm Mpc})^3$, and $\sigma_v = 300\,{\rm km/s}$. All future plots will perturb one feature of this. The left panel shows the true power spectrum (black), and the reconstructed power spectra using density-only (dashed), velocity-only (dotted), and joint density+velocity (solid). Three $\mu$ bins are considered, centered at $\mu=0.17$ (blue), $\mu=0.5$ (orange), and $\mu=0.83$ (green). The middle panel shows the cross-correlation between the reconstructed and true fields, while the right panel shows the transfer function between the two.}
  \label{fig:fiducial}
\end{figure}

\begin{figure}
  \begin{center}
    \includegraphics[width=1\textwidth]{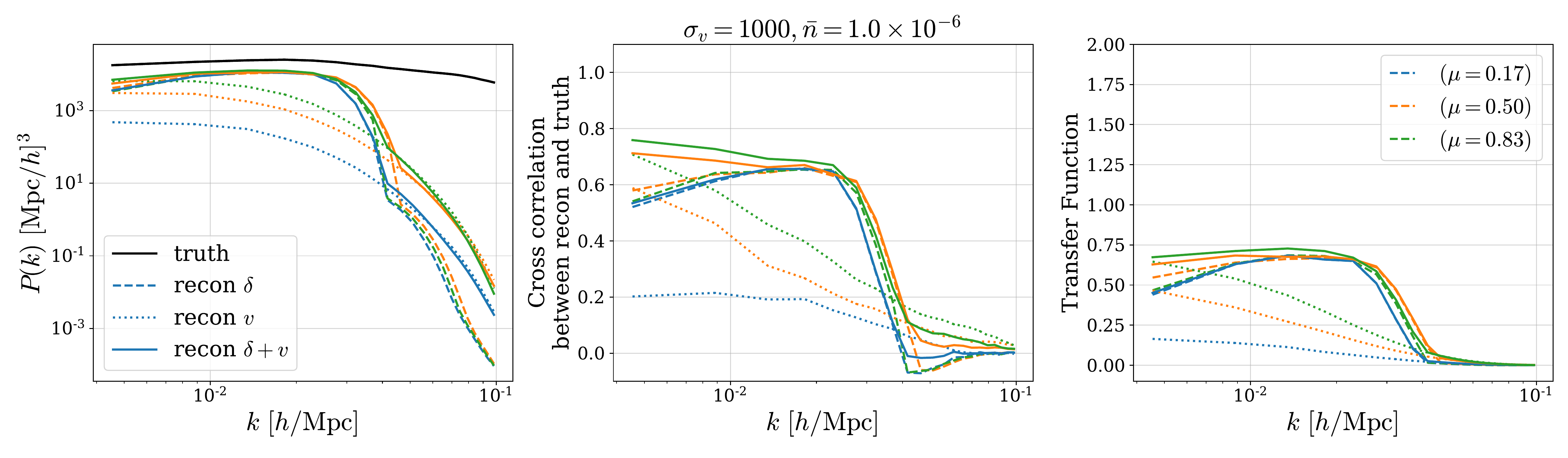}
  \end{center}
  \caption{Like top row of  Fig.~\ref{fig:fiducial} but with a higher velocity noise of $\sigma_{v,{\rm data}}=1000\,{\rm km/s}$.}
  \label{fig:high_vn}
\end{figure}

\begin{figure}
  \begin{center}
    \includegraphics[width=1\textwidth]{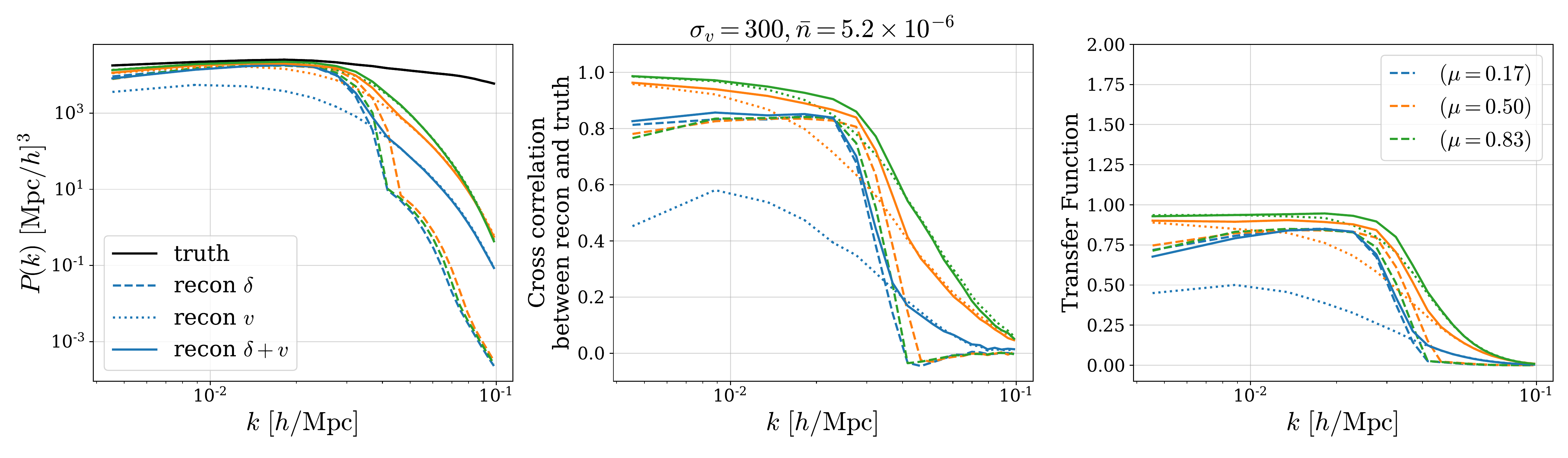}
  \end{center}
  \caption{Like top row of Fig.~\ref{fig:fiducial} but with a $\approx$5 times higher number density of $\bar{n}=5.2\times10^{-6}\,(h/{\rm Mpc})^3$.}
  \label{fig:high_Nh}
\end{figure}

\begin{figure}
  \begin{center}
    \includegraphics[width=1\textwidth]{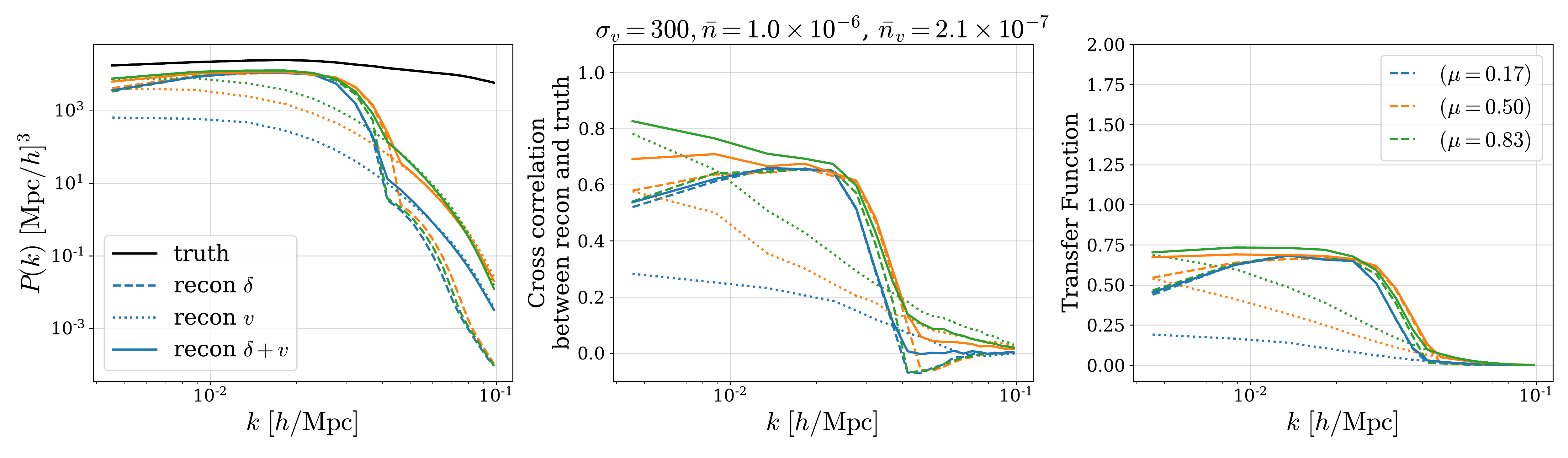}
  \end{center}
  \caption{Like top row of Fig.~\ref{fig:fiducial} but with $\approx$5 times fewer halos with peculiar velocity data.}
  \label{fig:low_nbarv}
\end{figure}

\begin{figure}
\vspace{-1cm}
  \begin{center}
    \includegraphics[width=\textwidth]{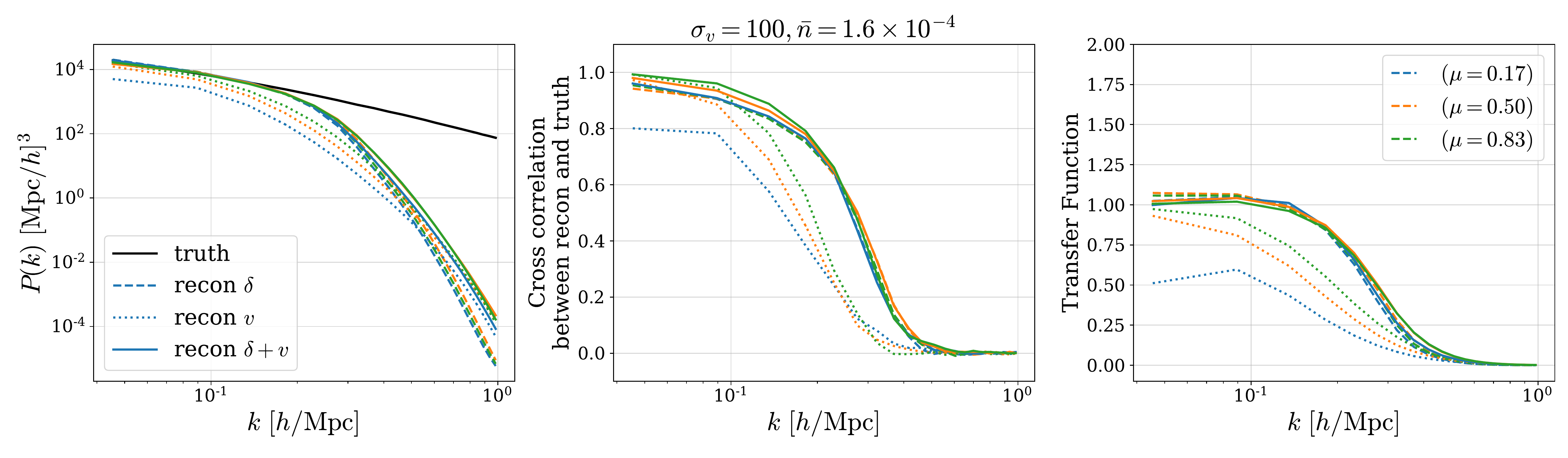}
    \includegraphics[width=\textwidth]{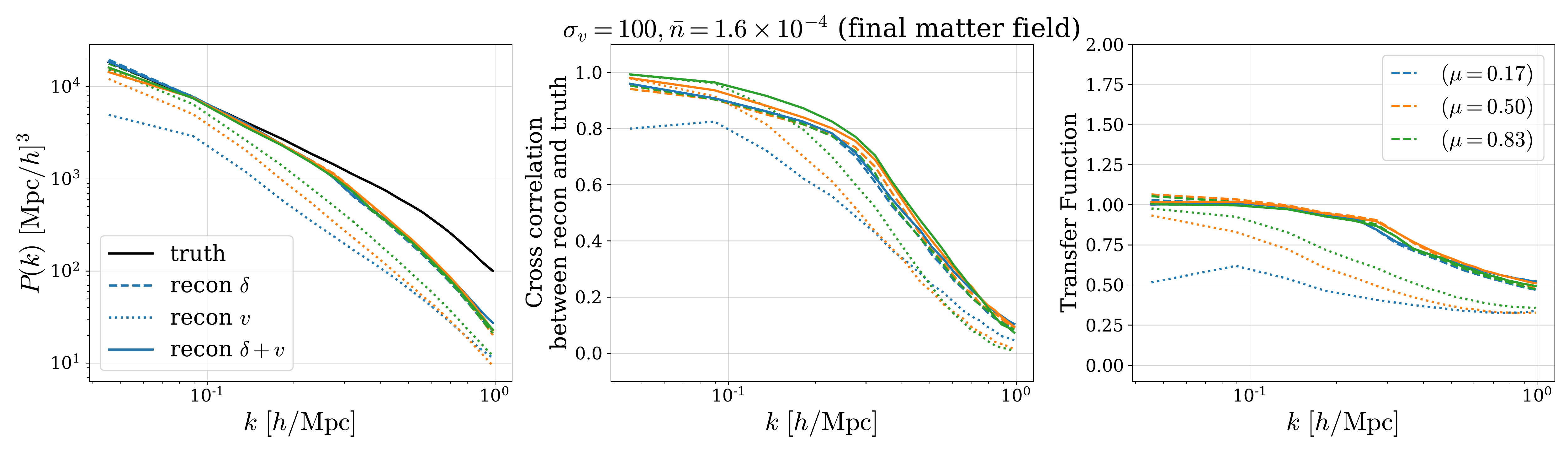}
    \includegraphics[width=\textwidth]{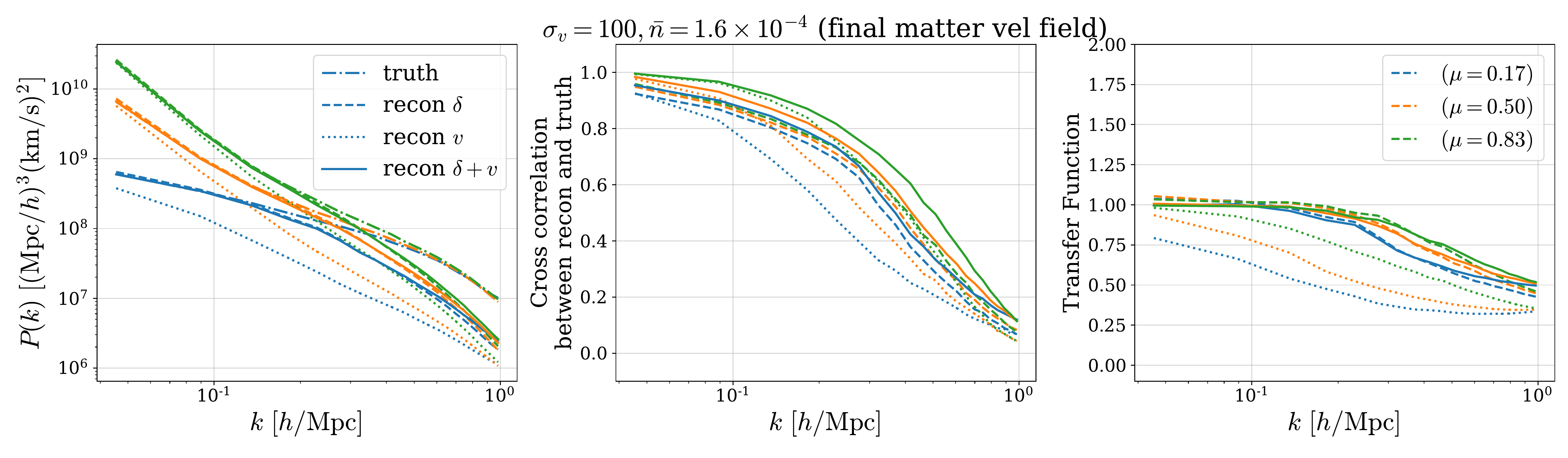}
  \end{center}
  \caption{Like Fig.~\ref{fig:fiducial} but with 10 times smaller box size (number of halos different now due to different resolution of simulation), and $\sigma_v=100 {\rm km/s}$.}
  \label{fig:low_bs}
\end{figure} 

We quantify the reconstruction accuracy in terms of the cross-correlation coefficient and transfer function between the reconstructed field and the data field. 
Because we consider only line-of-sight velocity data, we plot the reconstruction quality in three equally sized $\mu$ bins (with centers $\mu=0.17,0.5,0.83$), where $\mu\equiv \cos \theta$ and $\theta$ is the angle to the line of sight for an infinitely far away observer (corresponding to the flat-sky approximation). Given the power spectrum of the data field $P_{\rm data}$, the power spectrum of the reconstructed field $P_{\rm recon}$, and the cross power spectrum $P_{\rm data, recon}$, the cross-correlation coefficient is given by
\begin{equation}
    r(k, \mu) = \frac{P_{{\rm data, recon}}(k,\mu)}{\sqrt{P_{\rm data}(k,\mu)P_{\rm recon}(k,\mu)}},
\end{equation}
and the transfer function is given by
\begin{equation}
    T(k,\mu) = \sqrt{\frac{P_{\rm recon}(k,\mu)}{P_{\rm data}(k,\mu)}}.
\end{equation}
We note that the power spectrum in a ($k,\mu$) bin is calculated as 
\begin{equation}
    P(k,\mu) = \frac{1}{N_{k,\mu}V} \Sum_{\bi{k} ~:~ |\bi{k}|=k,~\bi{k}\cdot\bi{k}_z=\mu} \delta(\bi{k})\delta(-\bi{k}),
\end{equation}
where the sum runs over all wavevectors with magnitude $k$ (plus/minus the bin width) and angle to line of site $\mu$ (plus/minus the bin width), $N_{k,\mu}$ is the number of modes in the ($k,\mu$) bin, and $V$ is the volume of the box.

Fig.~\ref{fig:fiducial} shows the reconstruction for the fiducial setup described in Section \ref{sec:data}, namely: $4\, {\rm Gpc}/h$,  $\bar{n}=10^{-6}\, (h/ {\rm Mpc})^3$, and $\sigma_v = 300\,{\rm km/s}$. 
We first focus on the top row which shows the reconstruction of the initial linear field.
The middle panel shows that the reconstruction using only density data (dashed) is decorrelated on large and intermediate scales with a correlation value of approximately 0.6. Furthermore, the right panel shows the transfer function to be approximately 0.5 on the largest scales.
The poor reconstruction on small scales is due to the annealing described in Sec.~\ref{sec:opt} -- as we only anneal to $k_{\rm max}=16 k_F$ smaller scales are never fitted.
On the other hand, doing velocity-only reconstruction (dotted) greatly improves the large-scale correlation in the two highest $\mu$ bins, while the $\mu=0.17$ performs worse than density-only reconstruction. The story is similar for the transfer function which is increased to 0.8 in the highest $\mu$ bin on large scales.
Finally we consider reconstruction using the joint density and velocity data, using the full posterior of Eqn.~\ref{eqn:posterior} (solid). It can be seen that this further improves the correlation and transfer function on intermediate scales.
The second and third rows of Fig.~\ref{fig:fiducial} show the reconstruction of the final matter density and velocity fields respectively. In both cases the reconstruction quality is of similar quality to the initial field.

Fig.~\ref{fig:high_vn} considers increasing the error in the velocity data from 300 to 1,000 km/s. In this case the density reconstruction is unchanged compared to the fiducial. The velocity-only reconstruction now performs worse than in the fiducial setup and is only slightly better than density-only for both the correlation coefficient and transfer function along the line of sight. Combining velocity with density slightly improves upon density-only reconstruction along the line of sight.

Fig.~\ref{fig:high_Nh} considers increasing the number of halos in the fiducial setup by a factor of approximately 5 by lowering the minimum mass cut; the number density is now $\bar{n}=5.2\times10^{-5}\, (h/ {\rm Mpc})^3$ and the minimum halo mass $M_{\rm min}\simeq1.2 \times 10^{15} M_\odot / h$. Given the higher number density, there is now lower shot noise and density-only reconstruction performs better compared to the fiducial case.
Velocity-only reconstruction also benefits from the inclusion of additional halos, as there is now 5 times more halo data to use for reconstruction. The correlation coefficient and transfer function are now unity on the largest scales in the two highest $\mu$ bins.

So far we have assumed all halos used for density reconstruction also have velocity data. This is the case for kSZ, however, galaxy surveys are typically only able to measure the velocities of the most massive halos.
Fig.~\ref{fig:low_nbarv} considers the modification of the fiducial setup such that only the most massive 20\% of halos have velocity data. It can be seen that the velocity, and thus joint, reconstruction performs worse, with the correlation coefficient and transfer function dropping to around 0.8 on the largest scales for largest $\mu$ bin. 

Until now we have considered velocity reconstruction on large scales in the high shot noise regime. To study the effect on smaller scales we divide the box size of the fiducial setup by a factor of 10, giving a $400\,{\rm Mpc}/h$ box. We also consider a lower velocity noise of $\sigma_{v, {\rm data}}=100 \,{\rm km/s}$. The top row of Fig.~\ref{fig:low_bs} shows the reconstruction of the initial linear field. It can be seen that there is a small gain from joint reconstruction compared to density-only in terms of the correlation coefficient and transfer function.
We also plot the reconstruction of the final (nonlinear) matter density and peculiar velocity fields in the second and third rows of Fig.~\ref{fig:low_bs}. The reconstruction of the final density can be seen to be of comparable quality to the initial field reconstruction. One the other hand, the reconstruction of the velocity field is improved on small scales by performing joint reconstruction compared to using density or velocity alone.

\section{Discussion and Conclusions}
\label{sec:conclusions}

In this paper we have developed the formalism for including peculiar velocity information in field-level reconstruction of the initial conditions of the Universe. We have implemented it in the differentiable forward modelling code \texttt{FlowPM} to reconstruct the initial conditions using halo overdensity data, halo peculiar velocity data, and a combination of the two. We also considered the reconstruction of the final matter 
density and velocity fields. We showed that in shot noise dominated cases, reconstruction from density data alone is decorrelated from the truth, but this is greatly improved by including velocity data.

We studied this as a function of shot noise, error on velocity, and number of velocity tracers. We found that the benefit of including velocity data is very much dependent on these quantities.
We also showed that even in cases of low shot noise, combining velocity and density data, together with the non-linear model implemented in \texttt{FlowPM}, benefits the reconstruction of the final matter velocity field on nonlinear scales ($k > 0.1 \, h/{\rm Mpc}$).

We expect this work to have wide applicability to future surveys: for example, upcoming observations by DESI \cite{DESI_2016}, together with CMB maps from the Simons Observatory \cite{SO2018}, will produce kSZ measurements with signal-to-noise $\gtrsim 100$: indeed the reconstruction from velocities is expected to dominate in statistical power for scales $k \lesssim 0.02h$/Mpc \cite{Smith_2018} and be even more powerful with CMB-S4 \cite{CMB-S42016}. In addition, Rubin Observatory's LSST will discover hundreds of thousands of type Ia supernovae, for which a redshift can be obtained by DESI and individual peculiar velocities can be obtained with a few percent scatter \cite{Kim_2020}. In each case, forecasts based on linear theory suggest large improvements for measurement of cosmological parameters such as growth rate $f$ \cite{Kim_2020} and local primordial non-Gaussianity $f_{\rm NL}$ \cite{Munchmeyer_2019}. This is in part due to the lower noise overall in the reconstruction of the initial conditions, and in part to the fact that galaxy positions and velocities trace the same underlying matter density, and therefore quantities like $f$ and $f_{\rm NL}$ can be measured with reduced sample variance when combining the two measurements \cite{Seljak:2008xr}. The joint reconstruction formalism developed in this work is a natural way to optimally combine the data available, and inference of cosmological parameters from this is an important next step, which is left to future work.
To enable such future work, our code will be made publicly available upon publication of the paper.

\section*{Acknowledgements}
We thank Boryana Hadzhiyska, Alex Kim, Uro\v{s} Seljak, and Kendrick Smith for very useful discussion. SF is funded by the Physics Division of Lawrence Berkeley National Laboratory and by the U.S. Department of Energy (DOE), Office of Science, under contract
DE-AC02-05CH11231. This research used resources of the National Energy Research Scientific Computing Center, which is also supported by the Office of Science of the U.S.~DOE under Contract No.~DE-AC02-05CH11231.

\appendix
\section*{Appendices}
\section{Redshift Space Distortions}
\label{app:rsd}

Here we repeat the analysis performed in the main paper for the fiducial example, but include redshift space distortions in the data and the modeling. Redshift space distortions include some velocity information and so it is appropriate to see what effect this has relative to the peculiar velocity information.

To transform the data to redshift space, assuming the flat sky approximation, we shift the configurations space coordinate of each halo by $v_z / (aH)$, where $v_z$ is the velocity in the $z$ direction. 
To model redshift space distortions, we make two alterations to the model in the main paper. Firstly, while we use the same bias model as in the main paper, we now fit the bias parameters to the \textit{redshift space} halo field (see \cite{Schmittfull_2021} for application of the bias model to redshift space, note in particularly that the form of the model is the same to linear order, which is the order used in this work). Secondly, we displace the matter particles by their redshift space displacement, $v_z / (aH)$, before applying the bias model.

\begin{figure}
  \begin{center}
    \includegraphics[width=1\textwidth]{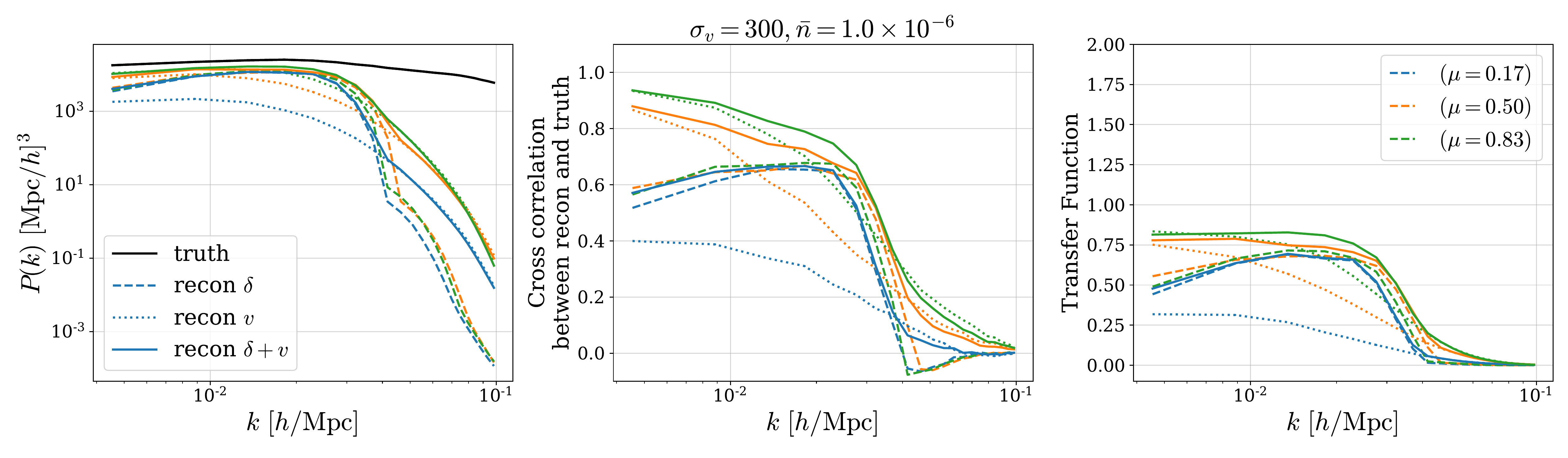}
  \end{center}
  \caption{Like top row of Fig.~\ref{fig:fiducial} but with redshift space distortions.}
  \label{fig:rsd}
\end{figure}

The results including redshift space distortions are shown in Fig.~\ref{fig:rsd}. By comparison to its configuration space analog, found in the top row of Fig.~\ref{fig:fiducial}, it can be seen that the quality of reconstruction is comparable. This is to be expected for halos on the scales considered in this work, however redshift space distortions will become more important when considering lower mass objects.

\bibliography{references}
\bibliographystyle{JHEP}

\end{document}